\documentclass[final,5p,times]{elsarticle}
\usepackage{amsmath}

\usepackage{caption}
\usepackage[labelfont=bf]{caption}
\usepackage[labelsep=period]{caption}
\journal{
}

\begin{document}

\begin{frontmatter}

\title{Insights into the structural symmetry of single-crystal YCrO$_3$ from \\ synchrotron X-ray diffraction}

\author[1,2,3]{Qian Zhao\fnref{fn1}}
\author[1,3]{Si Wu\fnref{fn1}}
\author[1]{Yinghao Zhu\fnref{fn1}}
\fntext[fn1]{These authors contributed equally.}
\author[1]{Junchao Xia}
\author[1]{Hai-Feng Li\corref{cor1}}
\cortext[cor1]{Corresponding author}
\ead{haifengli@um.edu.mo}
\affiliation[1]{organization={Joint Key Laboratory of the Ministry of Education, Institute of Applied Physics and Materials Engineering, University of Macau},
            addressline={Avenida da Universidade, Taipa},
            city={Macao SAR},
            postcode={999078},
            country={China}}
\affiliation[2]{organization={State Key Laboratory of High Performance Ceramics and Superfine Microstructure, Shanghai Institute of Ceramics, Chinese Academy of Science},
            city={Shanghai},
            postcode={200050},
            country={China}}
\affiliation[3]{organization={Guangdong--Hong Kong--Macao Joint Laboratory for Neutron Scattering Science and Technology},
            addressline={No. 1. Zhongziyuan Road, Dalang},
            city={DongGuan},
            postcode={523803},
            country={China}}

\begin{abstract}
We report on the crystallographic information such as lattice parameters, atomic positions, bond lengths and angles, and local crystalline distortion size and mode of single-crystal YCrO$_3$ compound by a high-resolution synchrotron X-ray diffraction study. The data was collected at 120 K (below $T_\textrm{N} \sim$ 141.5 K), 300 K (within [$T_\textrm{N}$, $T_\textrm{C}$]), and 500 K (above $T_\textrm{C} \sim$ 473 K). Taking advantages of high intensity and brilliance of synchrotron X-rays, we are able to refine collected patterns with the noncentrosymmetric monoclinic structural model ($P12_11$, No. 4) that was proposed previously but detailed structural parameters have not determined yet. Meanwhile, we calculated patterns with the centrosymmetric orthorhombic space group (\emph{Pmnb}, No. 62) for a controlled study. Lattice constants \emph{a}, \emph{b}, and \emph{c} as well as unit-cell volume almost increase linearly upon warming. We observed more dispersive distributions of bond length and angle and local distortion strength with the $P12_11$ space group. This indicates that (i) The local distortion mode of Cr2O$_6$ at 120 K correlates the formation of the canted antiferromagnetic order by Cr1-Cr2 spin interactions mainly through intermediate of O3 and O4 ions. (ii) The strain-balanced Cr1-O3(O4) and Cr2-O5(O5) bonds as well as the local distortion modes of Cr1O$_6$ and Cr2O$_6$ octohedra at 300 K may be a microscopic origin of the previously-reported dielectric anomaly. Our study demonstrates that local crystalline distortion is a key factor for the formation of ferroelectric order and provides a complete set of crystallography for a full understanding of the interesting magnetic and quasi-ferroelectric properties of YCrO$_3$ compound.


\end{abstract}




\end{frontmatter}


\section{Introduction}
\label{}

Biferroic materials, simultaneously accommodating both ferromagnetic and ferroelectric orders, are of particular interest due to potential couplings between the degrees of freedom as well as the resulting multiple applications \cite{wadhawan2000introduction, eerenstein2006multiferroic, martin2016thin, uchino2018ferroelectric}. Meanwhile, there exist great challenges for completely solving the actual crystalline structure toward a full understanding of the microscopic mechanisms of interesting macroscopic properties \cite{bertaut1966some, kim2007specific, alvarez2010weak, gervacio2018multiferroic, zhu2020high, zhu2020crystalline}. Theoretically, ferromagnetism and ferroelectricity appear to be mutually repulsion in oxides with perovskite-type structure \cite{wadhawan2000introduction, eerenstein2006multiferroic}. Therefore, such materials holding both ferromagnetic and ferroelectric degrees of freedom are few \cite{martin2016thin}. There exist many unsolved problems and reported controversies, for example, the disharmony between reported centrosymmetric structural model and appearance of ferroelectricity \cite{bertaut1966some, kim2007specific, alvarez2010weak, gervacio2018multiferroic, zhu2020high, zhu2020crystalline}. As a consequence, to date, there is still subject of numerous experimental and theoretical studies \cite{martin2016thin, uchino2018ferroelectric, zhu2020high, zhu2020crystalline, zhu2020enhanced, nicola2005, mishra2021observation, fita2021pressure, wang2021competition, gupta2020spin, sharma2021tuning, otsuka2021effect}.

The system of RECrO$_3$ (RE = rare earth) orthochromates has been attracting lots of attention. So far, their crystalline structure has been assigned as an orthorhombic one with centrosymmetric space group that holds an inversion charge center and cannot account for the appearance of ferroelectricity \cite{saha2014novel, zeng2021smoothing, zvezdin2021multiferroic}. Among RECrO$_3$ orthochromates, the YCrO$_3$ compound was believed to be a biferroic material displaying simultaneously weak ferromagnetism below $\sim$ 140 K and ferroelectric-like behaviors below $\sim$ 473 K \cite{Serrao2005}, together with potentials as catalyst \cite{poplawski2000catalytic}, negative-temperature-coefficient thermistor \cite{kim2003formation}, solid oxide fuel cell \cite{jabbar2017chromate}, and non-volatile memory application \cite{sharma2014unipolar}.

\begin{figure*}[!t]
\centering
\includegraphics[width = 0.73\textwidth] {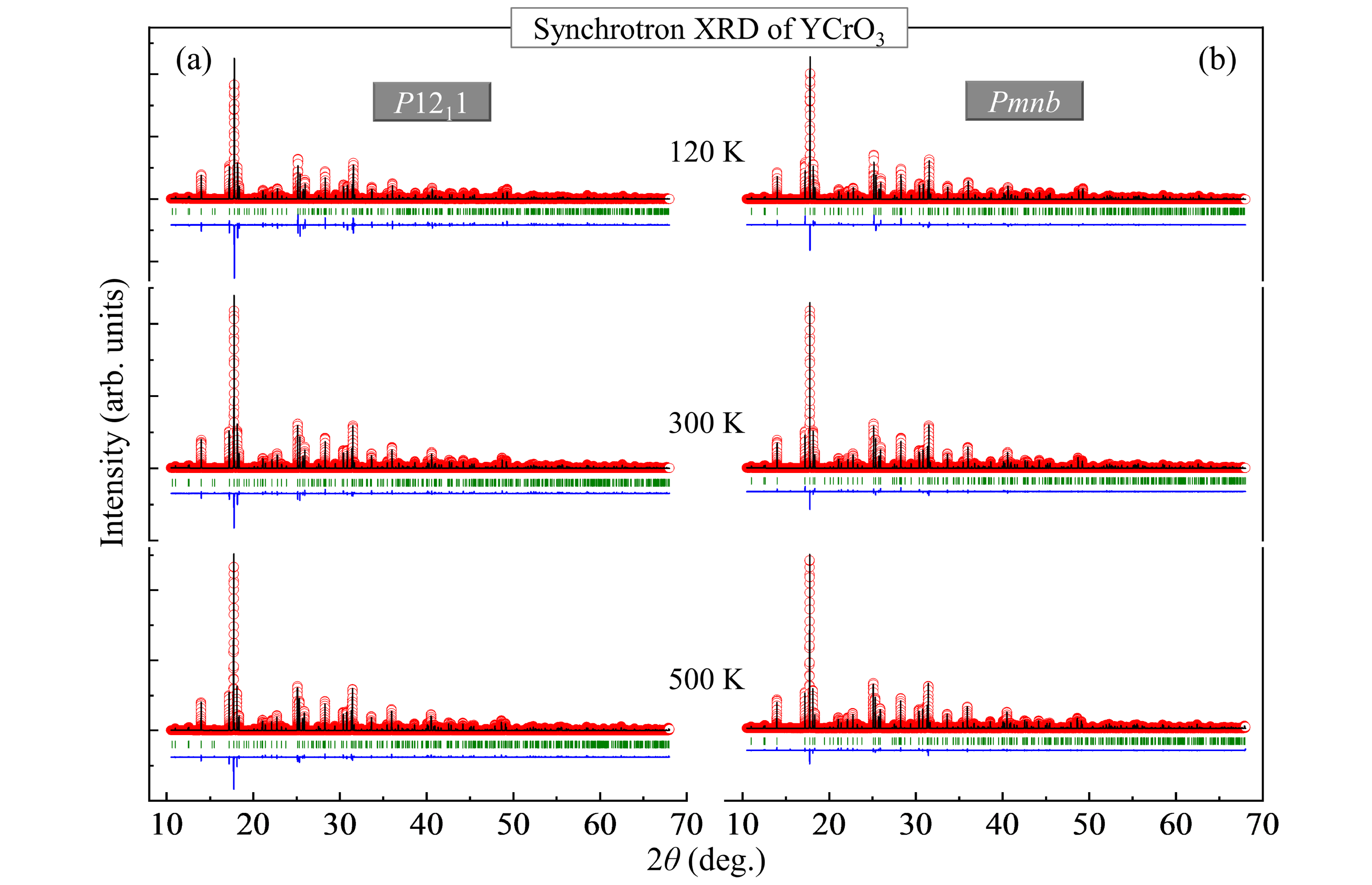}
\caption{
(a) Observed (circles) synchrotron X-ray powder diffraction patterns of a pulverized YCrO$_3$ single crystal, collected at 120 K (top panel), 300 K (middle panel), and 500 K (bottom panel). The solid lines are the calculated patterns with a noncentrosymmetric space group of $P12_11$ (No. 4). (b) The same synchrotron X-ray powder diffraction patterns as shown in (a) but were refined with a centrosymmetric space group of \emph{Pmnb} (No. 62). The vertical bars mark the positions of Bragg reflections. The lower curves represent the difference between observed and calculated patterns.
}
\label{SXRPD}
\end{figure*}

\begin{figure*}[!h]
\centering
\includegraphics[width = 0.73\textwidth] {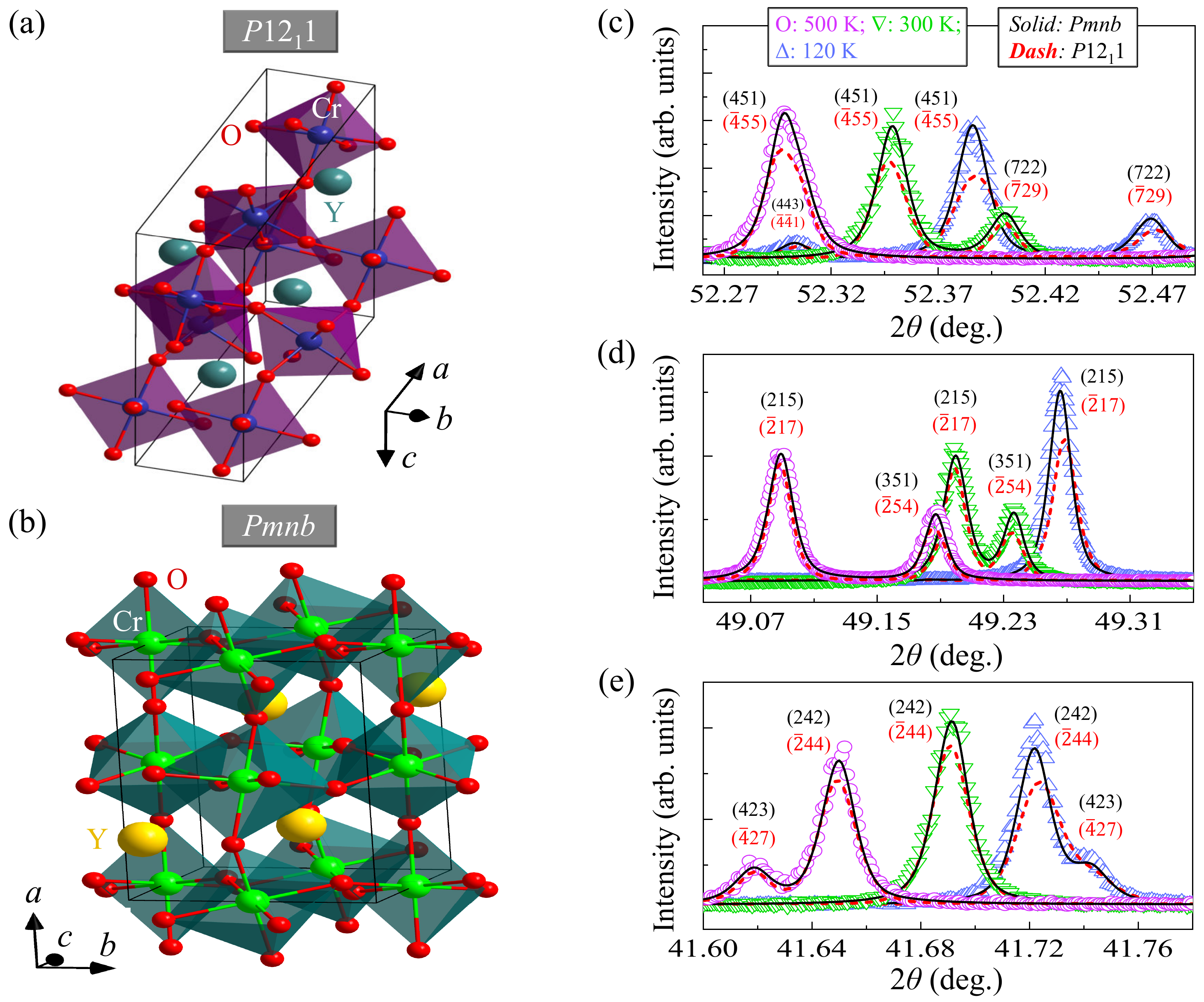}
\caption{
(a) Monoclinic crystal structure (space group: $P12_11$) with one unit cell (solid lines) of a YCrO$_3$ single crystal at 120 K. The octahedra of CrO$_6$ are highlighted in purple.
(b) Orthorhombic crystal structure (space group: $Pmnb$) with one unit cell (solid lines) of a YCrO$_3$ single crystal at 120 K. The octahedra of CrO$_6$ are highlighted in blue.
(c-e) For a clear comparison, we show vertically-shifted synchrotron X-ray powder diffraction patterns of a pulverized YCrO$_3$ single crystal at three representative 2$\theta$ angle regimes: 52.26--52.49$^{\circ}$ (c), 49.04--49.35$^{\circ}$ (d), and 41.60--41.78$^{\circ}$ (e), measured at temperatures of 120 K (up triangles), 300 K (down triangles), and 500 K (circles). The black solid lines are the calculations with \emph{Pmnb} space group (No. 62), and the red dashed lines are the calculations with $P12_11$ space group (No. 4). All main Bragg peaks were indexed with both space groups of \emph{Pmnb} (top black Miller indices as indicated) and $P12_11$ (bottom red Miller indices as marked).
}
\label{fit}
\end{figure*}

\begin{table*}[!t]
\renewcommand*{\thetable}{\Roman{table}}
\caption{Refined structural parameters of a pulverized YCrO$_3$ single crystal at 120, 300, and 500 K with space groups of $P12_11$ (No. 4) and \emph{Pmnb} (No. 62), including lattice constants, $\beta$ angle, unit-cell volume, atomic positions, and goodness of fit. We listed the Wyckoff site of each ion. The numbers in parenthesis are the estimated standard deviations of the (next) last significant digit.}
\label{RefinParas1}
\setlength{\tabcolsep}{2.2mm}{}
\renewcommand{\arraystretch}{1.1}
\begin{tabular}{llllllllll}
\hline
\multicolumn {10}{c}{A pulverized YCrO$_3$ single crystal: Controlled refinements with two space groups}                                                                                \\
\hline
Space group                                   &&\multicolumn{3}{c}{$P12_11$}                         &&\multicolumn{4}{c}{$Pmnb$}                                                       \\
$T$ (K)                                       &&120             &300                &500             &&                             &120             &300              &500             \\
\hline
$a$ ({\AA})                                   &&7.52424(1)      &7.53407(1)         &7.54756(1)      &&                             &7.52377(1)      &7.53350(1)       &7.54703(1)      \\
$b$ ({\AA})                                   &&5.52091(1)      &5.52315(1)         &5.52623(1)      &&                             &5.52045(1)      &5.52272(1)       &5.52583(1)      \\
$c$ ({\AA})                                   &&9.16631(2)      &9.17854(2)         &9.19557(2)      &&                             &5.23462(1)      &5.24183(1)       &5.25227(1)      \\
$\alpha (= \gamma)$ $(^\circ)$                &&90              &90                 &90              &&                             &90              &90               &90              \\
$\beta$ $(^\circ)$                            &&145.17181(8)    &145.17030(7)       &145.16501(6)    &&                             &90              &90               &90              \\
$V$ ({\AA}$^3$)                               &&217.467(1)      &218.138(1)         &219.086(1)      &&                             &217.418(1)      &218.088(1)       &219.039(1)      \\
\hline
Y1(2\emph{a}) \emph{x}                        &&0.2669(1)       &0.2670(1)          &0.2675(1)       &&Y1(4\emph{c}) \emph{x}       &0.25            &0.25             &0.25            \\
Y1(2\emph{a}) \emph{y}                        &&0.0673(1)       &0.0669(1)          &0.0661(1)       &&Y1(4\emph{c}) \emph{y}       &0.56757(3)      &0.56689(2)       &0.56608(3)      \\
Y1(2\emph{a}) \emph{z}                        &&0.2674(1)       &0.2670(1)          &0.2666(1)       &&Y1(4\emph{c}) \emph{z}       &0.48264(4)      &0.48299(3)       &0.48340(3)      \\
Y2(2\emph{a}) \emph{x}                        &&0.7331(1)       &0.7330(1)          &0.7325(1)       &&                             &                &                 &                \\
Y2(2\emph{a}) \emph{y}                        &&0.9327(1)       &0.9331(1)          &0.9340(1)       &&                             &                &                 &                \\
Y2(2\emph{a}) \emph{z}                        &&0.2326(1)       &0.2330(1)          &0.2334(1)       &&                             &                &                 &                \\
Cr1(2\emph{a}):                               &&                &                   &                &&Cr1(4\emph{b}):              &                &                 &                \\
$(x, y, z)$                                   &&(0, 0, 0.75)    &(0, 0, 0.75)       &(0, 0, 0.75)    &&$(x, y, z)$                  &(0, 0, 0.5)     &(0, 0, 0.5)      &(0, 0, 0.5)     \\
Cr2(2\emph{a}):                               &&                &                   &                &&                             &                &                 &                \\
$(x, y, z)$                                   &&(0.5, 0, 0.75)  &(0.5, 0, 0.75)     &(0.5, 0, 0.75)  &&                             &                &                 &                \\
O1(2\emph{a}) \emph{x}                        &&0.1427(1)       &0.1473(1)          &0.1505(1)       &&O1(4\emph{c}) \emph{x}       &0.25            &0.25             &0.25            \\
O1(2\emph{a}) \emph{y}                        &&0.4680(1)       &0.4663(1)          &0.4647(1)       &&O1(4\emph{c}) \emph{y}       &0.96643(27)     &0.96632(22)      &0.96519(22)     \\
O1(2\emph{a}) \emph{z}                        &&0.1470(1)       &0.1473(1)          &0.1476(1)       &&O1(4\emph{c}) \emph{z}       &0.60153(27)     &0.60268(22)      &0.60192(22)     \\
O2(2\emph{a}) \emph{x}                        &&0.8573(1)       &0.8527(1)          &0.8495(1)       &&                             &                &                 &                \\
O2(2\emph{a}) \emph{y}                        &&0.5320(1)       &0.5337(1)          &0.5353(1)       &&                             &                &                 &                \\
O2(2\emph{a}) \emph{z}                        &&0.3530(1)       &0.3527(1)          &0.3524(1)       &&                             &                &                 &                \\
O3(2\emph{a}) \emph{x}                        &&0.3664(1)       &0.3603(1)          &0.3537(1)       &&O2(8\emph{d}) \emph{x}       &0.05380(15)     &0.05378(11)      &0.05375(11)     \\
O3(2\emph{a}) \emph{y}                        &&0.3144(1)       &0.3030(1)          &0.2969(1)       &&O2(8\emph{d}) \emph{y}       &0.30362(20)     &0.30297(15)      &0.30204(16)     \\
O3(2\emph{a}) \emph{z}                        &&0.5689(1)       &0.5565(1)          &0.5474(1)       &&O2(8\emph{d}) \emph{z}       &0.30675(20)     &0.30654(16)      &0.30671(16)     \\
O4(2\emph{a}) \emph{x}                        &&0.6336(1)       &0.6397(1)          &0.6463(1)       &&                             &                &                 &                \\
O4(2\emph{a}) \emph{y}                        &&0.6856(1)       &0.6970(1)          &0.7031(1)       &&                             &                &                 &                \\
O4(2\emph{a}) \emph{z}                        &&0.9311(1)       &0.9435(1)          &0.9526(1)       &&                             &                &                 &                \\
O5(2\emph{a}) \emph{x}                        &&0.7359(1)       &0.7527(1)          &0.7646(1)       &&                             &                &                 &                \\
O5(2\emph{a}) \emph{y}                        &&0.2034(1)       &0.1970(1)          &0.1930(1)       &&                             &                &                 &                \\
O5(2\emph{a}) \emph{z}                        &&0.0457(1)       &0.0565(1)          &0.0658(1)       &&                             &                &                 &                \\
O6(2\emph{a}) \emph{x}                        &&0.2641(1)       &0.2473(1)          &0.2354(1)       &&                             &                &                 &                \\
O6(2\emph{a}) \emph{y}                        &&0.7966(1)       &0.8030(1)          &0.8070(1)       &&                             &                &                 &                \\
O6(2\emph{a}) \emph{z}                        &&0.4543(1)       &0.4435(1)          &0.4342(1)       &&                             &                &                 &                \\
\hline
$R_\textrm{p}$                                &&19.4            &16.2               &14.4            &&                             &12.5            &8.95             &7.93            \\
$R_\textrm{wp}$                               &&25.5            &21.6               &19.4            &&                             &15.3            &11.4             &11.4            \\
$R_\textrm{exp}$                              &&6.60            &6.57               &6.45            &&                             &6.87            &6.55             &6.40            \\
$\chi^2$                                      &&14.9            &10.9               &9.02            &&                             &4.96            &3.02             &3.16            \\
\hline
\end{tabular}
\end{table*}

\begin{table*}[!t]
\renewcommand*{\thetable}{\Roman{table}}
\caption{Extracted bond lengths, bond angles, and the distortion parameter $\Delta$ of a pulverized YCrO$_3$ single crystal at 120, 300, and 500 K with $P12_11$ (No. 4) and \emph{Pmnb} (No. 62) space groups, respectively. The numbers in parenthesis are the estimated standard deviations of the (next) last significant digit.}
\label{RefinParas2}
\setlength{\tabcolsep}{1.8mm}{}
\renewcommand{\arraystretch}{1.1}
\begin{tabular}{llllllll}
\hline
\multicolumn {8}{c}{A pulverized YCrO$_3$ single crystal: Controlled refinements with two space groups}                                                                                        \\
\hline
Space group                                   &                &$P12_11$           &                  &                               &                &$Pmnb$           &                     \\
$T$ (K)                                       &120             &300                &500               &                               &120             &300              &500                  \\
\hline
Cr1-O1(O2) (\AA)                              &1.9338(1)       &1.9678(1)          &1.9927(1)         &Cr1-O1 (\AA)                   &1.9606(4)       &1.9676(3)        &1.9706(3)            \\
Cr1-O3(O4) (\AA)                              &1.9912(1)       &1.9824(1)          &1.9699(1)         &Cr1-O21 (\AA)                  &2.0000(11)      &1.9980(9)        &1.9952(9)            \\
Cr1-O5(O6) (\AA)                              &2.0069(1)       &1.9985(1)          &1.9909(1)         &Cr1-O22 (\AA)                  &1.9795(11)      &1.9824(9)        &1.9890(9)            \\
$<$Cr1-O$>$ (\AA)                             &1.9773(1)       &1.9829(1)          &1.9845(1)         &$<$Cr1-O$>$  (\AA)             &1.9800(3)       &1.9827(3)        &1.9849(3)            \\
Cr2-O1(O2) (\AA)                              &1.9958(1)       &1.9678(1)          &1.9508(1)         &                               &                &                 &                     \\
Cr2-O3(O4) (\AA)                              &2.0098(1)       &1.9985(1)          &2.0013(1)         &                               &                &                 &                     \\
Cr2-O5(O6) (\AA)                              &1.9646(1)       &1.9824(1)          &2.0096(1)         &                               &                &                 &                     \\
$<$Cr2-O$>$ (\AA)                             &1.9901(1)       &1.9829(1)          &1.9872(1)         &                               &                &                 &                     \\
\hline
$\angle$Cr1-O1(O2)-Cr2 $(^\circ)$             &146.420(1)      &146.346(1)         &146.259(1)        &$\angle$Cr1-O1-Cr1 $(^\circ)$  &147.223(1)      &146.346(1)       &146.452(1)           \\
$\angle$Cr1-O3(O4)-Cr2 $(^\circ)$             &143.902(1)      &146.052(1)         &147.458(1)        &$\angle$Cr1-O2-Cr1 $(^\circ)$  &145.824(59)     &146.054(47)      &146.170(47)          \\
$\angle$Cr1-O5(O6)-Cr2 $(^\circ)$             &146.608(1)      &146.052(1)         &144.694(1)        &                               &                &                 &                     \\
\hline
$\Delta$(Y1) $(\times 10^{-4})$               &48.159          &42.262             &44.718            &                               &40.724          &42.230           &42.603               \\
$\Delta$(Y2) $(\times 10^{-4})$               &48.159          &42.264             &44.785            &                               &                &                 &                     \\
$\Delta$(Cr1) $(\times 10^{-4})$              &2.522           &0.398              &0.269             &                               &0.664           &0.392            &0.275                \\
$\Delta$(Cr2) $(\times 10^{-4})$              &0.901           &0.402              &1.711             &                               &                &                 &                     \\
$\Delta$(O1) $(\times 10^{-4})$               &53.841          &50.868             &49.924            &                               &54.779          &50.881           &50.356               \\
$\Delta$(O2) $(\times 10^{-4})$               &53.840          &50.865             &50.000            &                               &                &                 &                     \\
$\Delta$(O3) $(\times 10^{-4})$               &125.483         &131.629            &140.423           &                               &130.085         &131.605          &132.797              \\
$\Delta$(O4) $(\times 10^{-4})$               &125.483         &131.630            &140.412           &                               &                &                 &                     \\
$\Delta$(O5) $(\times 10^{-4})$               &143.801         &131.601            &128.605           &                               &                &                 &                     \\
$\Delta$(O6) $(\times 10^{-4})$               &143.801         &131.603            &128.576           &                               &                &                 &                     \\
\hline
\end{tabular}
\end{table*}

\begin{figure}[t]
\centering
\includegraphics[width = 0.50\textwidth] {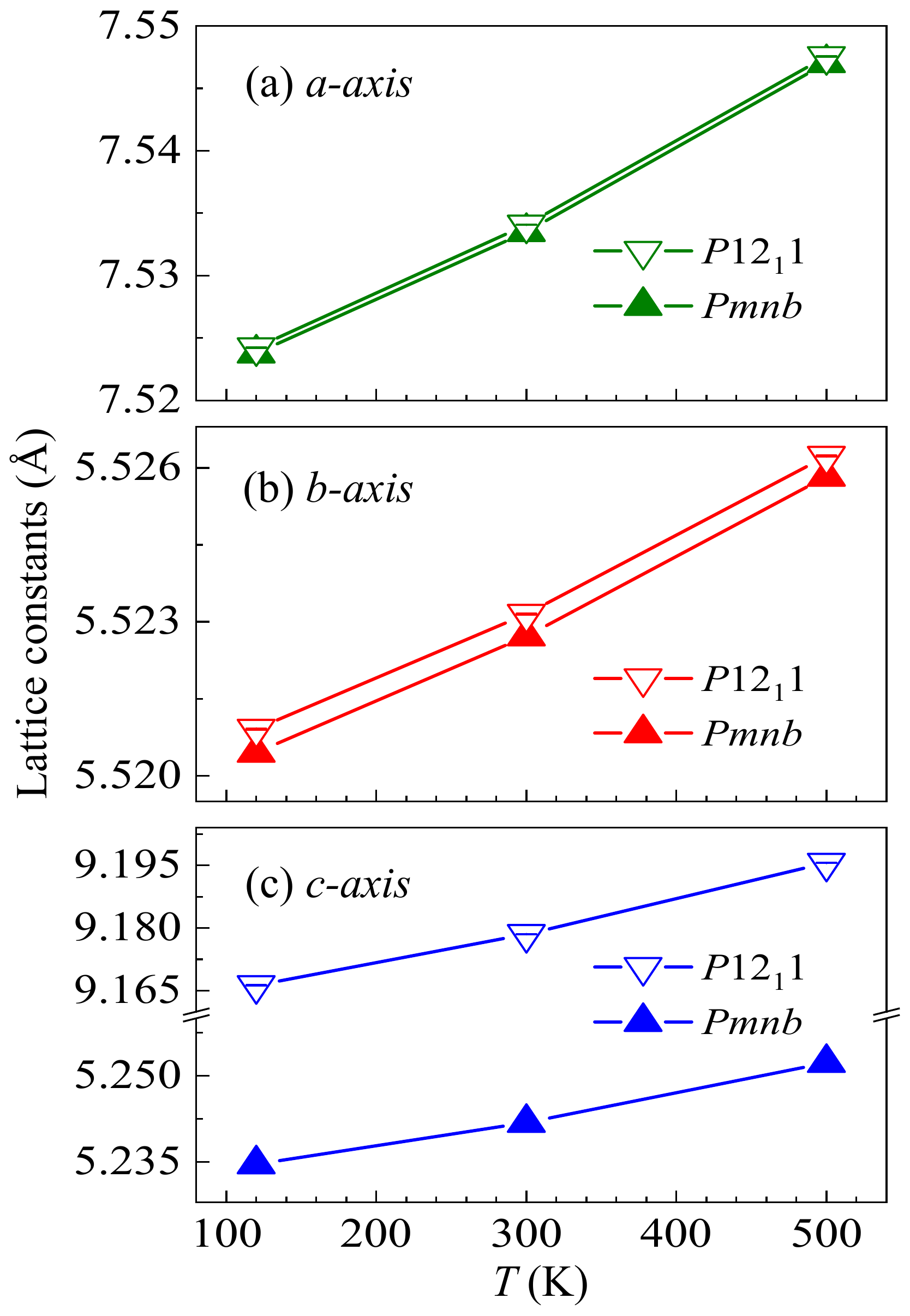}
\caption{
Temperature-dependent lattice constants \emph{a} (a), \emph{b} (b), and \emph{c} (c) of a pulverized YCrO$_3$ single crystal, which was extracted from our refinements with the two space groups of $P12_11$ (down void triangles) and $Pmnb$ (up solid triangles), based on the synchrotron X-ray powder diffraction data collected at 120, 300 and 500 K. The $P12_11$ space group belongs to a monoclinic structure so the length of the \emph{c} constant is greater than the corresponding one in orthorhombic $Pmnb$ structure. Error bars are the standard deviations obtained from our refinements. The solid lines are guides to the eye.
}
\label{latticeP}
\end{figure}

\begin{figure}[t]
\centering
\includegraphics[width = 0.50\textwidth] {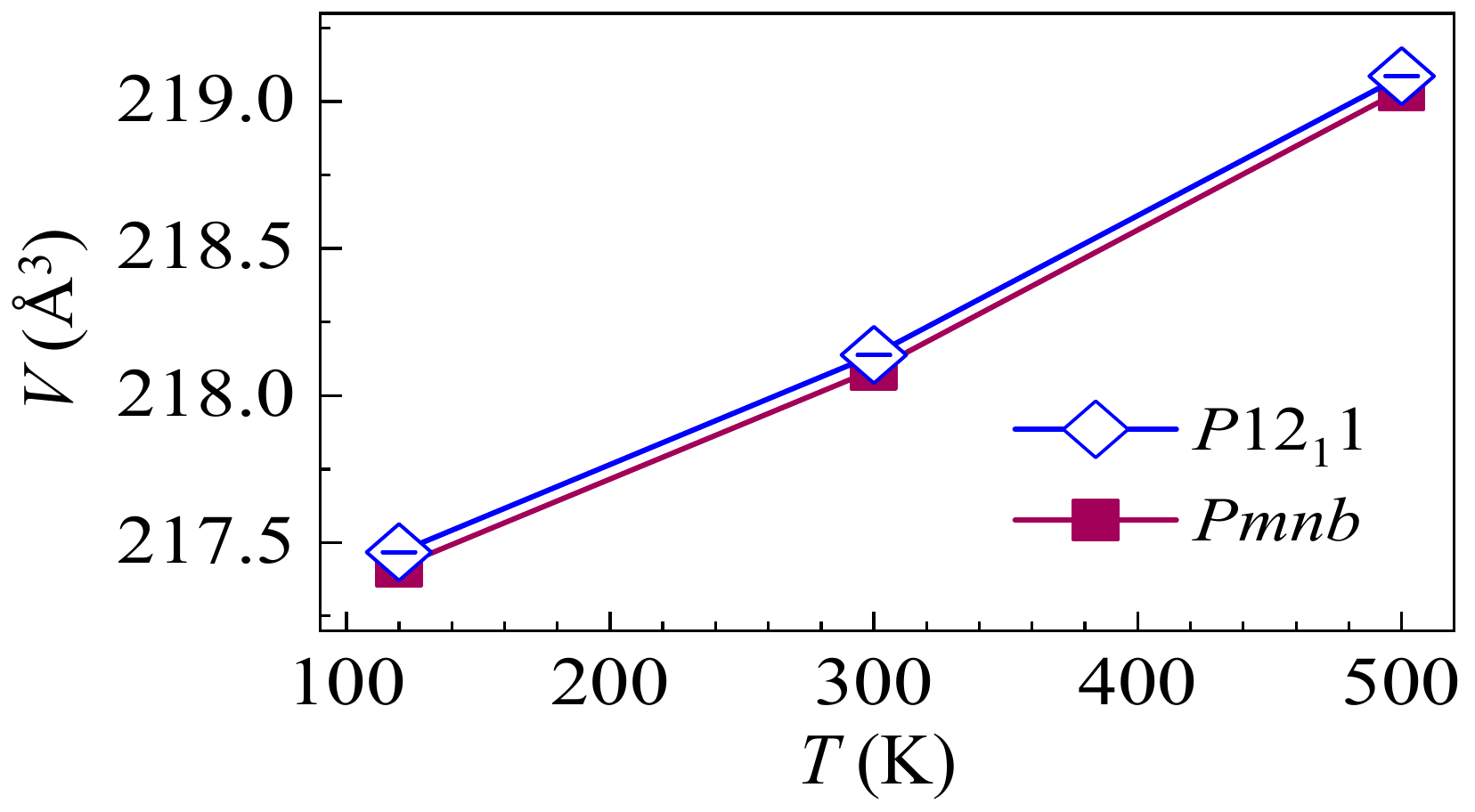}
\caption{
Temperature-dependent unit-cell volume, \emph{V}, extracted from our refinements with space groups of $P12_11$ (void rhombuses) and $Pmnb$ (solid rectangles). Error bars are the standard deviations obtained from our refinements. The solid lines are guides to the eye.
}
\label{UCVolume}
\end{figure}

\begin{figure*}[t]
\centering
\includegraphics[width = 0.73\textwidth] {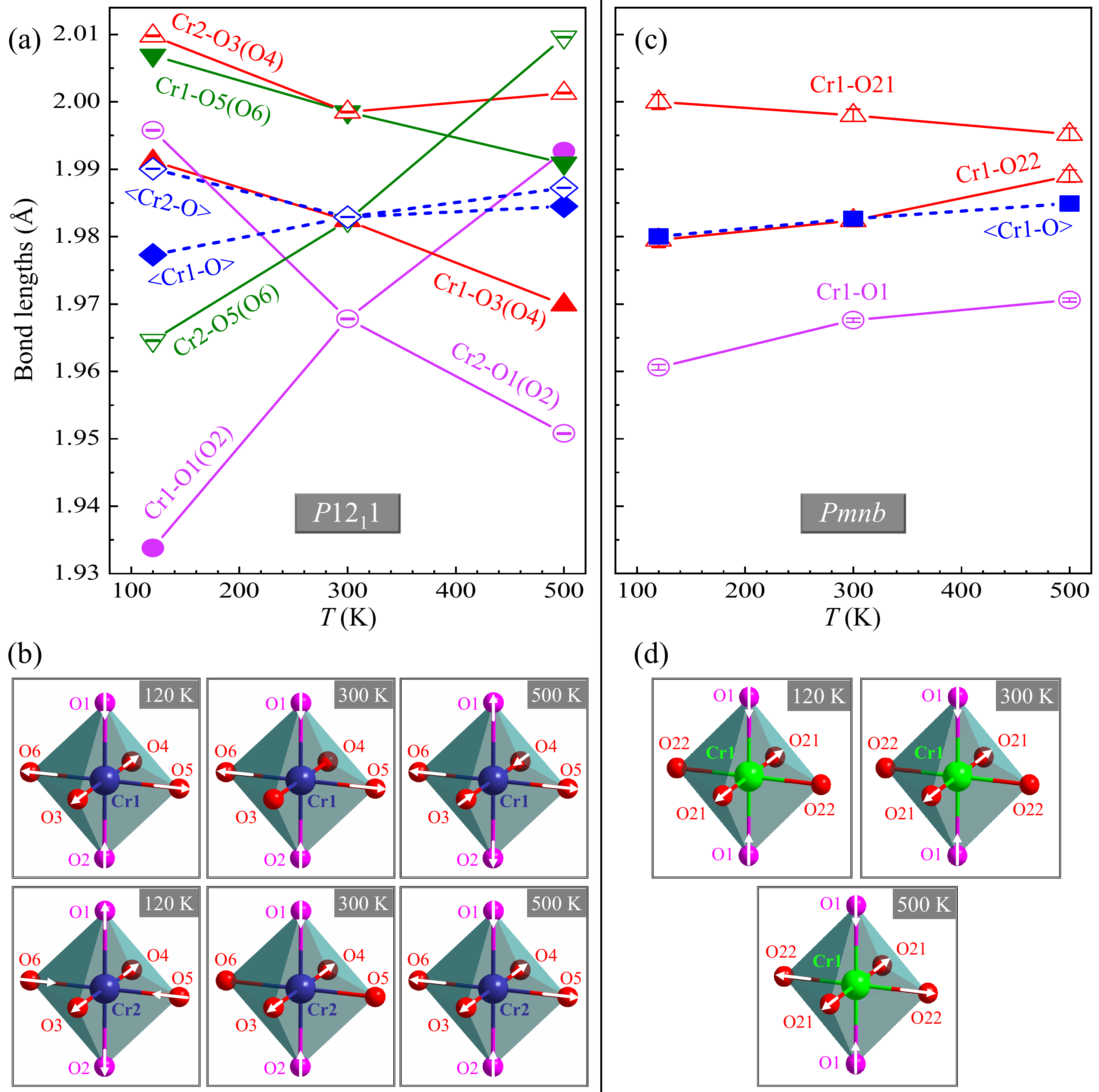}
\caption{
(a) Temperature-dependent bond lengths of Cr1-O1(O2), Cr1-O3(O4), Cr1-O5(O6), Cr2-O1(O2), Cr2-O3(O4), and Cr2-O5(O6), as well as the averaged bond lengths of $<$Cr1-O$>$ and $<$Cr2-O$>$, in single-crystal YCrO$_3$ compound with $P12_11$ (No. 4) space group.
(b) The local octahedral environments of Cr1(Cr2) ions in $P12_11$ space group at 120, 300, and 500 K.
(c) Temperature-dependent bond lengths of Cr1-O1, Cr1-O21, and Cr1-O22, as well as the averaged bond length of $<$Cr1-O$>$, obtained with \emph{Pmnb} (No. 62) space group. Error bars in (a) and (c) are standard or propagated deviations.
(d) The local octahedral environments of Cr1 ions in \emph{Pmnb} space group at 120, 300, and 500 K.
In (b) and (d), arrows sitting on the O ions and Cr-O bonds schematically represent the deduced octahedral distortion modes.
}
\label{BondLth}
\end{figure*}

\begin{figure*}[h]
\centering
\includegraphics[width = 0.82\textwidth] {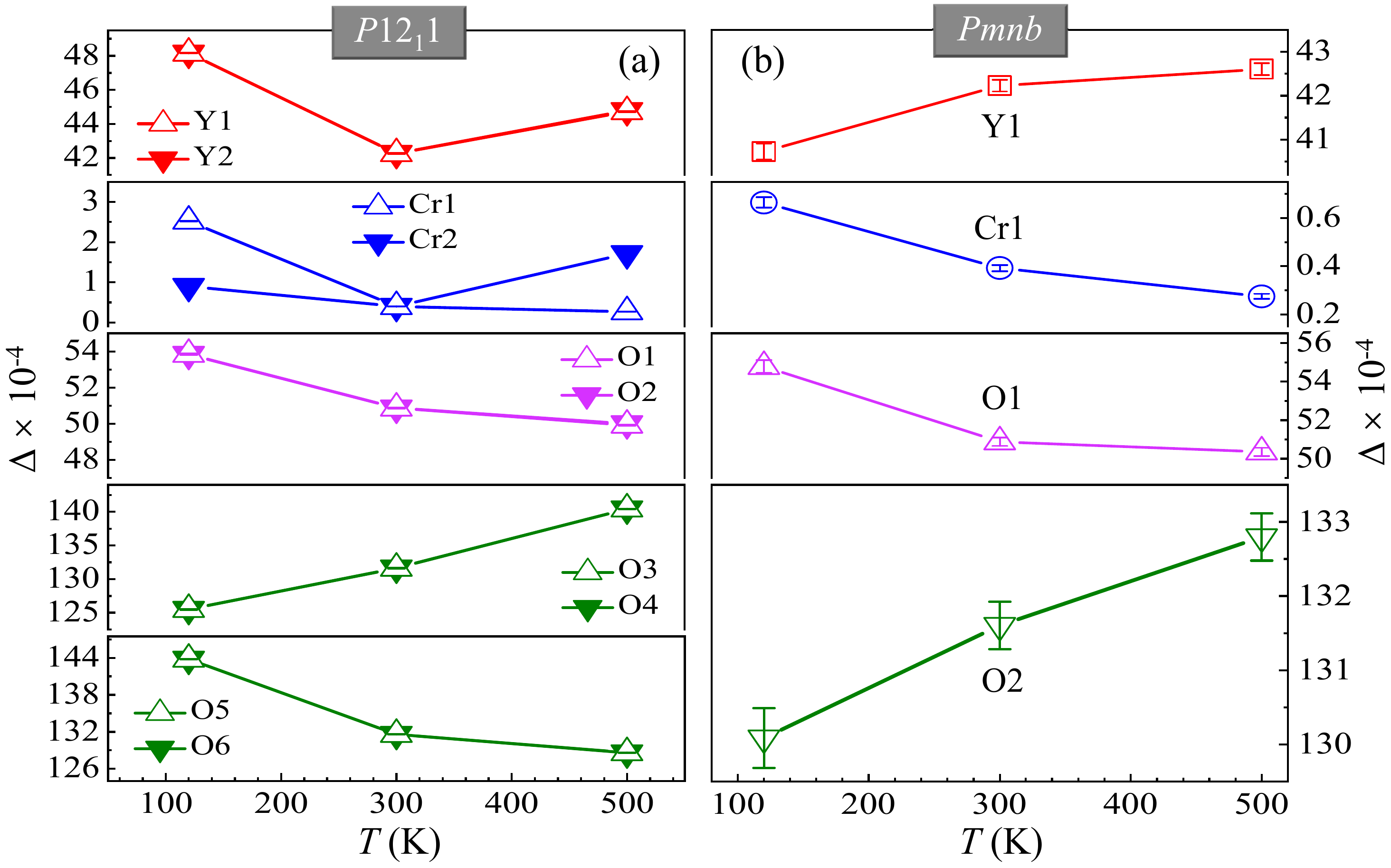}
\caption{
Temperature variation of the distortion parameter, $\Delta$, of a single-crystal YCrO$_3$, refined from the data measured at 120, 300, and 500 K with (a) space group $P12_11$ (No. 4) for Y1, Y2, Cr1, Cr2, O1, O2, O3, O4, O5, and O6 ions and (b) space group \emph{Pmnb} (No. 62) for Y1, Cr1, O1, and O2 ions. Error bars are the standard deviations extracted from our refinements. The solid lines are guides to the eye.
}
\label{distortion}
\end{figure*}

\begin{figure*}[!h]
\centering
\includegraphics[width = 0.73\textwidth] {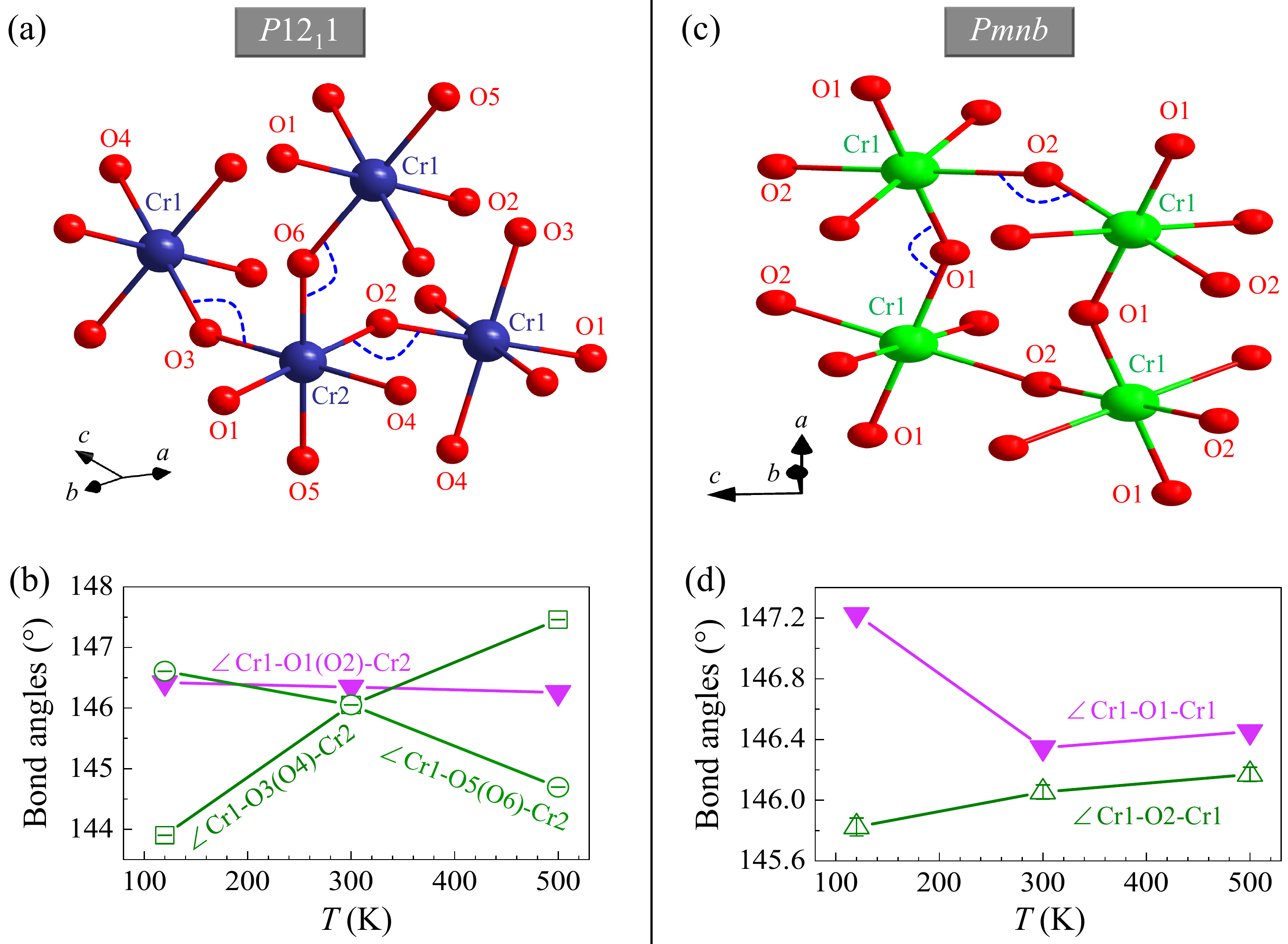}
\caption{
In the structure with space group of $P12_11$, illustrations of the bond angles of Cr1-O1(O2)-Cr2, Cr1-O3(O4)-Cr2, and Cr1-O5(O6)-Cr2 (a) and their temperature variations (b). In the structure with space group of \emph{Pmnb}, illustrations of the bond angles of Cr1-O1-Cr1 and Cr1-O2-Cr1 (c) and their temperature dependencies (d).
}
\label{BondAng}
\end{figure*}

Up to now, most of the previous studies indicate that YCrO$_3$ belongs to an orthorhombic structure with space group of \emph{Pnma} \cite{bertaut1966some, kim2007specific, gervacio2018multiferroic, sahu2008modification, duran2010magneto, ardit2010elastic, sardar2011direct, singh2013anomalous, krishnan2013synthesis, jara2018optical, sanina2018electric, sharma2020epitaxial, chakraborty2021structural}, for example, a high-resolution neutron powder diffraction study \cite{ramesha2007observation} revealed that below and above $\sim$ 473 K at which the dielectric anomaly occurs \cite{Serrao2005}, the average crystal structure of YCrO$_3$ compound is orthorhombic with space group of $Pnma$ (No. 62). Such an orthorhombic structure is centrosymmetric, which makes it not easy to form Dzyaloshinskii–Moriya interactions and ferroelectric polarizations. Furthermore, a monoclinic structure with space group of $P2_1/n$ and lattice parameters of $a$ $\approx$ $c$ = 7.61 $\pm$ 0.01 {\AA}, $b$ = 7.54 $\pm$ 0.01 {\AA}, and $\beta$ = 92$^{\circ}$56$'$ $\pm$ 6$'$ were determined for YCrO$_3$ \cite{looby1954yttrium, katz1955unit}. The excessive phonon entropy \cite{kim2007specific} was attributed to possible monoclinic structure ($P2_1/n$, No. 14, centrosymmetric) \cite{looby1954yttrium, katz1955unit} of YCrO$_3$, and the additional Debye phonon modes in heat capacity data \cite{kim2007specific} were suspected to be related to the space group of $P12_11$ (No. 4, noncentrosymmetric) \cite{Serrao2005}. Moreover, first-principles calculations indicated that the lowest energy structure of YCrO$_3$ was a monoclinic system with the noncentrosymmetric space group of $P12_11$ \cite{Serrao2005}. A neutron study on YCrO$_3$ with pair distribution function (PDF) method \cite{ramesha2007observation} demonstrates that over the range of 1.6--6 {\AA}, the high-temperature (550 K) data can be well indexed with the space group of $Pnma$, by contrast, the data collected in the ferroelectric state could be better fit with the noncentrosymmetric $P12_11$ model, leading to a Cr off-centring displacement of the order of 0.01 {\AA} along the $z$ direction. Therefore, local noncentrosymmetry was suggested to be the origin of the ferroelectricity in YCrO$_3$ \cite{Serrao2005}. It is pointed out that so far, the theoretically and experimentally proposed monoclinic ($P12_11$) model is the only noncentrosymmetric one \cite{Serrao2005}, reconciling the existence of ferroelectricity in YCrO$_3$.

As is well known, the PDF study \cite{ramesha2007observation} provides information of short-range or local crystalline structures from the diffuse scattering. The monoclinic ($P12_11$) structure is one of the subgroups of the orthorhombic ($Pnma$) one. Within the space group of $Pnma$, YCrO$_3$ accommodates Y1, Cr1, O1, and O2 atomic sites, while in $P12_11$ they transfer into Y1/Y2, Cr1/Cr2, O1/O2, and O3/O4/O5/O6 sites, respectively, making a more complicated crystalline structure. It is stressed that so far, it remains an unsolved issue to determine the lattice parameters and atomic positions of the monoclinic ($P12_11$) model \cite{zhu2020high, zhu2020crystalline, ramesha2007observation, zhao2021temperature}.

In this paper, to reconcile the contradiction between the previously-observed dielectric anomaly \cite{Serrao2005} and the present consensus on a centrosymmetric crystal structure \cite{sanina2018electric, sharma2020epitaxial, chakraborty2021structural, ramesha2007observation} of YCrO$_3$ compound, we carried out a high-resolution synchrotron X-ray powder diffraction on a pulverized YCrO$_3$ single crystal. The collected data was refined with two space groups of centrosymmetric orthorhombic \emph{Pmnb} (No. 62) and noncentrosymmetric monoclinic $P12_11$ (No. 4), trying to identify the real crystalline structure and symmetry. From values of goodness of fit, it is evident that the orthorhombic model is more suitable for the average macroscopic crystalline structure of YCrO$_3$ compound in consistent with the arrived comment consent. On the other hand, our study also confirms that the $P12_11$ structure extracted by the neutron PDF study \cite{ramesha2007observation} indeed exists locally in YCrO$_3$ compound. Taking advantage of the synchrotron X-ray powder diffraction, we are able to refine the collected high-quality data with $P12_11$ model and determine the corresponding lattice parameters, atomic positions, bond lengths and angles, and local distortion modes, addressing the existing issue on crystallographic information of $P12_11$ model and providing a basis for further experimental and theoretical studies. Meanwhile, we also extracted the crystallographic information of \emph{Pmnb} model with the same data for a controlled study.

\section{Experimental}

We prepared polycrystalline YCrO$_3$ samples by solid state reaction \cite{li2008synthesis} with a procedure similar to the previously-reported one \cite{zhu2020high, zhu2020crystalline}. The YCrO$_3$ single crystals were grown with a laser-diode floating-zone furnace (Model: LD-FZ-5-200W-VPO-PC-UM) well-equipped at the University of Macau, Macao, China \cite{wu2020super, li2021patent}.

We carefully chose and gently ground a YCrO$_3$ single crystal into powdered sample with a Vibratory Micro Mill (FRITSCH PULVERISETTE 0) for the synchrotron X-ray powder diffraction study to determine temperature-dependent crystallographic information \cite{li2014magnetization}. The measurements were performed on the beamline I11 at Diamond Light Source, Didcot, UK \cite{thompson2009beamline, parker2011high, tang2015design}. X-rays with a wavelength of $\lambda =$ 0.827 {\AA} were used as the radiation source. High-resolution powder diffraction patterns were collected over a diffraction 2$\theta$ angle range of 10--70$^\circ$ at 120, 300, and 500 K. The 2$\theta$ step interval is 0.001$^{\circ}$, and the counting time is 1800 s. The YCrO$_3$ powder was loaded onto the outside surface of a 0.3 mm diameter borosilicate glass capillary tube by attaching a thin layer of hand cream. The capillary sample holder fit directly to a magnetic spinner and was mounted at the center of the $\theta$ circle face plate.

We used the computer program FULLPROF SUITE \cite{fullprof} to analyze all collected data. The peak profile shape was modeled with a Pseudo-Voigt function. We refined the background with a linear interpolation between automatically-selected background points. We refined the scale factor, zero shift, peak shape parameters, asymmetry, lattice constants, and atomic positions, etc.

\section{Results and discussion}

\subsection{High-resolution synchrotron X-ray powder diffraction}

In order to clarify the structural effect on the magnetic and ferroelectric properties of single-crystal YCrO$_3$ compound, we carried out synchrotron X-ray powder diffraction measurements at 120 K (below the antiferromagnetic transition temperature $T_\textrm{N}$ $\sim$ 141.5 K \cite{zhu2020crystalline}), 300 K (between $T_\textrm{N}$ and $T_\textrm{C}$), and 500 K (above the ferroelectric transition temperature $T_\textrm{C}$ $\sim$ 473 K \cite{Serrao2005}) to probe the actual crystallographic symmetry. Figure~\ref{SXRPD} shows the collected synchrotron X-ray powder diffraction patterns as well as the corresponding structural refinements with space groups of $P12_11$ (Fig.~\ref{SXRPD}(a)) and \emph{Pmnb} (Fig.~\ref{SXRPD}(b)). It is clear that the collected data can be well indexed with both space groups. The extracted unit cells were illustrated in Figs.~\ref{fit}(a) ($P12_11$) and~\ref{fit}(b) (\emph{Pmnb}). The refined crystallographic information such as lattice parameters and atomic positions was listed in Table~\ref{RefinParas1}, and the calculated bond lengths and angles as well as the local distortion parameter $\Delta$ were recorded into Table~\ref{RefinParas2}. From a comparison between the values of goodness of fit (Table~\ref{RefinParas1}), we found that the values of $R_\textrm{p}$, $R_\textrm{wp}$, $R_\textrm{exp}$, and $\chi^2$ from refinements with space group of \emph{Pmnb} are much lower than the corresponding values from refinements with space group of $P12_11$, for example, $\chi^2 =$ 3.02 (\emph{Pmnb}) and 10.9 ($P12_11$) for the data collected at 300 K. In principle, with increasing the number of refinable parameters while transferring space group from \emph{Pmnb} into $P12_11$, the goodness of fit should become better \cite{fullprof, li2009crystal}, i.e., the goodness of fit gets lower values. Our refinements derogate from the above principle, which on the contrary demonstrates that the proper average macroscopic structural model for the YCrO$_3$ single crystal is \emph{Pmnb} rather than $P12_11$ and indicates that the $P12_11$ structure merely exists locally in YCrO$_3$.

We carefully checked temperature evolution of the shape and position of Bragg peaks as marked in Figs.~\ref{fit}(c-e) within three representative 2$\theta$ regimes. As a whole, the refinements with \emph{Pmnb} space group (shown as solid lines in Figs.~\ref{fit}(c-e)) fit better to the Bragg peaks. The Bragg peak (451) observed in \emph{Pmnb} space symmetry at 500 K evolved into (451) and (722) peaks at 300 K and (443), (451), and (722) reflections at 120 K, and the corresponding peak position shifted to higher angles (Fig.~\ref{fit}(c)); the observed (215) and (351) Bragg peaks at 500 and 300 K combined into one (215) peak at 120 K (Fig.~\ref{fit}(d)); the Bragg (423) and (242) peaks appeared at 500 K first united into (242) peak at 300 K and then split into (242) and (423) peaks again (Fig.~\ref{fit}(e)). The shift of Bragg positions indicates changes in lattice constants.

Traditionally, powder diffraction including X-ray and neutron is usually utilized to solve crystalline structures \cite{David2006}. Especially, neutron scattering is a powerful technique to solve magnetic structures \cite{xiao2010neutron} and monitor spin excitations and fluctuations \cite{PhysRevB.82.140503}. Compared to neutron sources, although synchrotron X-ray has much stronger intensity and brilliance \cite{David2006}, we did not observe any magnetic indications for single-crystal YCrO$_3$ at 120 K (Fig.~\ref{SXRPD}).

\subsection{Temperature dependent lattice parameters}

Figure~\ref{latticeP} shows the refined lattice constants \emph{a} (Fig.~\ref{latticeP}(a)), \emph{b} (Fig.~\ref{latticeP}(b)), and \emph{c} (Fig.~\ref{latticeP}(c)) of the YCrO$_3$ single crystal with space groups of $P12_11$ and \emph{Pmnb}. With increasing temperature from 120 to 500 K, the lattice constants \emph{a}, \emph{b}, and \emph{c} almost increase linearly. The values of lattice constants \emph{a} and \emph{b} in $P12_11$ space group are a little larger than the ones in space group of \emph{Pmnb}. It is pointed out that in $P12_11$ space group, the monoclinic (mc) lattice \textbf{c} axis is transformed from (-\textbf{a}-\textbf{c}) in orthorhombic (or) \emph{Pmnb} space group, which leads to the crystallographic angle $\beta \neq$ 90$^\circ$ (Table~\ref{RefinParas1}) and the value of constant $c_{\textrm{mc}} = \sqrt{a^2_{\textrm{or}} + c^2_{\textrm{or}}}$ (Fig.~\ref{latticeP}(c)). The temperature variances of lattice constants as well as the value change in $\beta$ angle jointly result in the change in unit-cell volumes with temperature as shown in Fig.~\ref{UCVolume}.

\subsection{Temperature dependent bond lengths and local distortion modes}

Local crystalline environments such as bond lengths and angles as well as their unhomogeneous distributions may have dramatic impacts on macroscopic properties of the host. Figure~\ref{BondLth} shows calculated bond lengths of Cr and O ions in the YCrO$_3$ single crystal with space groups of $P12_11$ (Figs.~\ref{BondLth}(a) and~\ref{BondLth}(b)) and \emph{Pmnb} (Figs.~\ref{BondLth}(c) and~\ref{BondLth}(d)).

We first discuss the results extracted with the space group of \emph{Pmnb} (Figs.~\ref{BondLth}(c) and~\ref{BondLth}(d)). With increasing temperature from 120 to 500 K, the bond lengths of Cr1-O1 and Cr1-O22 increase, whereas that of Cr1-O21 decreases, which jointly leads to an increase of the averaged bond length $<$Cr1-O$>$ (Fig.~\ref{BondLth}(c)). Based on these results, we extracted the local distortion modes of Cr1O$_6$ octahedra at different temperatures (Fig.~\ref{BondLth}(d)). At 500 K, i.e., in the paraelectric state, Cr1-O21 and Cr1-O22 bonds are stretched, and Cr1-O1 bonds are compressed, displaying like a cooperative Jahn-Teller distortion. Upon cooling, when $T =$ 300 K, YCrO$_3$ stays in the quasi-ferroelectric state, and the Cr1-O22 bonds get balanced, i.e., without any stretching and suppressing. With further cooling down to 120 K, although YCrO$_3$ undergoes an antiferromagnetic phase transition at $T_\textrm{N}$ $\sim$ 141.5 K, the Cr1O$_6$ local distortion mode keeps similar to that at 300 K. Therefore, the change in Cr1-O22 bonds responses sensitively to the observed dielectric anomaly.

Secondly, we present the results obtained from refinements with $P12_11$ space group (Figs.~\ref{BondLth}(a) and~\ref{BondLth}(b)). As shown in Fig.~\ref{BondLth}(a): (i) For the Cr1 ions, as temperature increases from 120 to 500 K, the values of Cr1-O3(O4) and Cr1-O5(O6) bond lengths decrease, by contrast, that of Cr1-O1(O2) bonds increases. (ii) For the Cr2 ions, Cr2-O1(O2) and Cr2-O3(O4) bond lengths decrease, whereas that of Cr2-O5(O6) increases. Fig.~\ref{BondLth}(b) shows the extracted local distortion modes of Cr1(Cr2)O$_6$ octahedra: (i) When $T =$ 500 K, the distortion mode of Cr2O$_{(1-6)}$ so obtained resembles that of Cr1O$_{(1,2)}$ in \emph{Pmnb} space group. For the Cr1O$_{(1-6)}$ octahedron, Cr1-O1(O2) and Cr1-O3(O4) bonds change their strained states, i.e., from the shortened to a prolongated state for Cr1-O1(O2) bonds and an opposite effect on Cr1-O3(O4) bonds. Cr1-O5(O6) bonds keep a similar state to that of Cr2-O5(O6). (ii) At 300 K, Cr1-O3(O4) and Cr2-O5(O6) bonds become normal, and other Cr1(Cr2)-O bonds keep their respective states as that of Cr2O$_6$ at 500 K. (iii) At 120 K, Cr1(Cr2)-O3(O4), Cr1-O5(O6), and Cr2-O1(O2) bonds are stretched, and Cr1-O1(O2) and Cr2-O5(O6) bonds are compressed.

By carefully comparing all the local distortion modes in $P12_11$ space group displayed in Fig.~\ref{BondLth}(b), it is clear that Cr1O$_6$ at 120 K displays the same local distortion mode as that of Cr2O$_6$ at 500 K, therefore, the local distortion mode of Cr2O$_6$ at 120 K is sensitive to the formation of the antiferromagnetic order. The local distortion modes of Cr1O$_6$ and Cr2O$_6$ octohedra at 300 K, beside the relaxation of Cr1-O3(O4) and Cr2-O5(O6) bonds, may be the key factors for a full understanding of the quasi-ferroelectric order.

\subsection{Temperature dependent strength of the local crystalline distortion}

The strength of the local crystalline distortion of CrO$_6$ octahedra can be estimated by the parameter \cite{li2006correlation, li2007neutron},
\begin{eqnarray}
\Delta = \frac{1}{n}\sum_{i=1}^{n}\left(\frac{d_{i}-\langle d\rangle}{\langle d\rangle}\right)^2,
\label{delta}
\end{eqnarray}
where \emph{n} is the coordination number, which equals six in a CrO$_6$ octahedral, \emph{d$_i$} is the bond length along one of the \emph{n} coordination directions, and $\langle d \rangle$ is the mean Cr--O bond length. As shown in Fig.~\ref{distortion} and Table~\ref{RefinParas2}, the local distortion parameter $\Delta$ extracted with \emph{Pmnb} space group has similar values as previously reported \cite{zhu2020crystalline, zhao2021temperature}: $\Delta$ of Cr1 ions keeps a lower value of $<$ 1 $\times$ 10$^{-4}$, while Y1, O1, and O2 ions have more than one order of magnitude larger $\Delta$ values, indicating that stronger local distortions exist in Y1, O1, and O2 ions. This could be a key to understand the dielectric anomaly \cite{Serrao2005, Mall2020}. As temperature increases, the $\Delta$ values of Y1 and O2 ions increase, whereas those of Cr1 and O1 ions decrease (Fig.~\ref{distortion}(b)). While refining the data with $P12_11$ space group, despite that Y1(Y2), O1(O2), O3(O4), and O5(O6) ions locate at different crystalline positions (Table~\ref{RefinParas1}), the values of their distortion parameter $\Delta$ are almost the same within error bars and display a decreasing tread upon warming except for the O3(O4) ions. It is hard to distinguish their responses to the antiferromagnetic phase transition and the quasi-ferroelectric transition. Meanwhile, the $\Delta$ value of Cr1 ions shows an increase at 120 K below the antiferromagnetic transition temperature, and that of Cr2 ions increases at 500 K above the quasi-ferroelectric transition temperature. This phenomenon implies that although Cr1 and Cr2 ions show the same behaviors in space group of \emph{Pmnb}, they are easily distinguishable in $P12_11$ space group and contribute differently to the formations of the antiferromagnetic and the quasi-ferroelectric phases.

\subsection{Temperature dependent bond angles}

Figure~\ref{BondAng} shows the space configurations and calculated values of Cr-O-Cr bond angles with space groups of $P12_11$ (Figs.~\ref{BondAng}(a) and~\ref{BondAng}(b)) and \emph{Pmnb} (Figs.~\ref{BondAng}(c) and~\ref{BondAng}(d)). Within the $P12_11$ crystal structure, $\angle$Cr1-O3(O4)-Cr2 is an angle along the intra-chain of CrO$_6$ octahedra, and $\angle$Cr1-O5(O6)-Cr2 is an inter-chain angle (Fig.~\ref{BondAng}(a)). The bond angles extracted from refinements with $P12_11$ space group show that the value of $\angle$Cr1-O3(O4)-Cr2 increases upon warming while that of $\angle$Cr1-O5(O6)-Cr2 decreases, and that of $\angle$Cr1-O1(O2)-Cr2 displays no obvious change (Fig.~\ref{BondAng}(b)). Thus thermal enhancements weaken the tilting of CrO$_6$ octahedra along the intra-chains and strengthen the rotating of CrO$_6$ octahedra along the inter-chains. As shown in Figs.~\ref{BondAng}(c) and ~\ref{BondAng}(d), the inter-chain angle $\angle$Cr1-O1-Cr1 shows a clear enhancement at 120 K, and the intra-chain angle $\angle$Cr1-O2-Cr1 displays a small decrease at 120 K in the \emph{Pmnb} space group.

It is well known that the value of bond angle $\angle$Cr-O-Cr has a strong effect on the strength of spin exchanges. Within the \emph{Pmnb} space group, magnetic interactions between Cr1 and Cr1 ions are stronger through the intermediate of O2 ions than via O1 ions while entering into the antiferromagnetic state upon cooling below $T_\textrm{N}$ (Fig.~\ref{BondAng}(d)). In the $P12_11$ space group, the spin interactions of Cr1-Cr2 ions occur mainly via O3(O4) ions (Fig.~\ref{BondAng}(b)).

\section{Conclusions}

In summary, we have collected high-resolution synchrotron X-ray powder diffraction data of single-crystal YCrO$_3$ compound at 120, 300, and 500 K, which was refined with two space groups of $P12_11$ (No. 4, noncentrosymmetric) and \emph{Pmnb} (No. 62, centrosymmetric). From physics point of view, the data can be well fit with both space groups. We extracted the structural parameters such as lattice constants, lattice angle $\beta$, unit-cell volume, atomic positions, and bond lengths and angles. We thus calculated the local crystalline distortion parameter $\Delta$. For the pairs of Y1 and Y2 ions, O1 and O2 ions, O3 and O4 ions, and O5 and O6 ions, their $\Delta$ values are almost the same in the $P12_11$ space group. From the calculated values of bond lengths, we extracted the detailed CrO$_6$ distortion modes and discussed their effects on the formations of the antiferromagnetic order and the quasi-ferroelectric state. From the values of refined Cr-O-Cr bond angles, we discussed the pathway of spin interactions. The crystallograhpic information of single-crystal YCrO$_3$ obtained in this study with $P12_11$ space group provides for the first time a set of complete structural parameters of the noncentrosymmetric space group, which may shed light on the observed dielectric anomaly in YCrO$_3$ compound and be very important for further experimental and theoretical studies.

\section*{CRediT authorship contribution statement}

\textbf{Qian Zhao:} Methodology, Investigation, Validation, Writing - Original Draft, Visualization.
\textbf{Si Wu:}  Methodology, Investigation, Validation, Writing - Original Draft, Visualization.
\textbf{Yinghao Zhu:} Methodology, Investigation, Validation, Writing - Original Draft, Visualization.
\textbf{Junchao Xia:} Methodology, Software, Investigation, Validation, Visualization.
\textbf{Hai-Feng Li:} Conceptualization, Writing - Review \& Editing, Visualization, Supervision, Project administration, Funding acquisition.

\section*{Declaration of Competing Interest}

The authors declare that they have no known competing financial interests or personal relationships that could have appeared to influence the work reported in this paper.

\section*{Acknowledgments}

Authors acknowledge financial supports from the opening project of State Key Laboratory of High Performance Ceramics and Superfine Microstructure (Grant No. SKL201907SIC), Science and Technology Development Fund, Macao SAR (File Nos. 0051/2019/AFJ and 0090/2021/A2), Guangdong Basic and Applied Basic Research Foundation (Guangdong--Dongguan Joint Fund No. 2020B1515120025), University of Macau (\\ 
MYRG2020-00278-IAPME and EF030/IAPME-LHF/2021/GDSTIC), and Guangdong--Hong Kong--Macao Joint Laboratory for Neutron Scattering Science and Technology (Grant No. 2019B121205003).

\bibliographystyle{elsarticle-num}
\bibliography{YCOSXRD-MRB}

\end{document}